\documentclass[english,aps,twocolumn]{revtex4}
\usepackage[T1]{fontenc}
\usepackage[latin9]{inputenc}
\usepackage{amssymb}
\usepackage{graphicx}
\usepackage{esint}
\PassOptionsToPackage{normalem}{ulem}
\usepackage{ulem}

\makeatletter

\@ifundefined{textcolor}{}
{%
 \definecolor{BLACK}{gray}{0}
 \definecolor{WHITE}{gray}{1}
 \definecolor{RED}{rgb}{1,0,0}
 \definecolor{GREEN}{rgb}{0,1,0}
 \definecolor{BLUE}{rgb}{0,0,1}
 \definecolor{CYAN}{cmyk}{1,0,0,0}
 \definecolor{MAGENTA}{cmyk}{0,1,0,0}
 \definecolor{YELLOW}{cmyk}{0,0,1,0}
 }

\usepackage {url}
\renewcommand\[{\begin{equation}}
\renewcommand\]{\end{equation}}

\makeatother

\usepackage{babel}
\begin{document}

\title{Searching for Topological Degeneracy in the Hubbard Model \\
with Quantum Monte Carlo}

\author{Bryan K. Clark}

\affiliation{Station Q, Microsoft Research, Santa Barbara, CA 93106, USA\\
}
\begin{abstract}
$Z_{2}$ spin liquids have topological order. One manifestation of
this is that a $Z_{2}$ spin liquid on a torus exhibits a four-fold
degeneracy.  Recent numerical evidence has argued for the existence
of a spin liquid ground state in the Hubbard model on a honeycomb
lattice near $U\approx4$ \cite{meng2010quantum}. The evidence for
this claim involves the presence of a gapped state that lacks any
identifiable order. This argument relies on being able to distinguish
small order from no order which is notoriously difficult. In this
paper we demonstrate an approach which uses quantum Monte Carlo to
search for one of the key features that positively identify the topological
spin liquid: the topological degeneracy. For any finite system, this
topological degeneracy is split where the splitting decays exponentially
with system size. We search for low lying states in the energy spectrum
that could be identified as these topologically degenerate states.
We show that, for system sizes up to $N=162$ sites, there is no evidence
for these states being significantly below the first excited state
giving evidence against the existence of a topological phase. We discuss
the possible options for the Hubbard model on the honeycomb lattice
in the absence of such degeneracy. 
\end{abstract}
\maketitle
\textbf{Introduction }Most physical systems are either gapless or
order at low temperatures, spontaneously breaking a symmetry. Systems
which fail to do so are usually exotic and often demonstrate interesting
properties such as topological order. In fact, a famous theorem of
Hastings \cite{hastings-2005-70} states that systems with an odd
number of sites per unit cell and no broken symmetries must either
be gapless or topologically ordered. One manifestation of this topological
order is that the system has a ground state degeneracy which depends
on the topological manifold on which the system lives. Recent work
\cite{meng2010quantum} has given strong numerical evidence that the
Hubbard model on the honeycomb lattice at intermediate $U$ ($3.5\lessapprox U\lessapprox4.25$)
is gapped (both in the spin and charge sectors) and has no order/broken
symmetries. As $Z_{2}$ spin liquids are gapped and typically have
no broken symmetries, this numerical result has been taken as evidence
for a spin-liquid ground state on the Hubbard honeycomb model. $Z_{2}$
spin liquids also have topological order \cite{balents2010spin,lee2008high,wen2002quantum}.
Because the honeycomb lattice has two sites per unit cell, Hasting's
theorem is not directly applicable. This leaves open a key question:
is the system actually topological? 

Either a positive or negative answer to this question is intrinsically
interesting. Finding topological order would give the first positive
indication of the $Z_{2}$ spin liquid state and supply the key missing
piece of evidence in making the case for this phase. To date, all
evidence for the spin liquid has involved the heroic task of ruling
out all other possible ordered states. This requires extrapolating
many order parameters to the thermodynamic limit and distinguishing
a small value from zero. We propose that searching for the ground
state degeneracy may be a more robust qualitative feature that doesn't
require distinguishing small differences. 

On the other hand, finding no ground state degeneracy would rule out
the interpretation of the numerical results given in ref. \cite{meng2010quantum}.
This then would leave three alternative possibilities: (1) There is
some order in the system. (2) The system is actually gapless. (3)
The ground state is a gapped phase with no broken symmetries and is
not topological. This latter possibility must be a Mott insulator
as band insulators that don't break symmetries are forbidden on the
honeycomb lattice at half filling \cite{kimchi2012featureless}. This
would be particularly intriguing as even this Mott insulating case
is forbidden when there are an odd number of sites per unit cell.
Although featureless insulators have been discussed in ref. \cite{kimchi2012featureless},
no simple local fermionic Hamiltonian which supports (3) as a potential
ground state are currently known. In addition, finding such an example
would further emphasize the importance of establishing the topological
nature as a necessary step in identifying spin liquids. 

In this work, we compute properties of the low-lying spectrum of the
Hubbard model on the honeycomb lattice with the motivation of looking
for {}``degenerate'' ground states. We accomplish this by developing
an approach which combines quantum Monte Carlo calculations with analytical
bounds on how high-energy states can influence the low temperature
entropy. In the process, we develop evidence against the presence
of topological degeneracy. We do this by showing that, for finite-size
systems, the low-lying spectrum of the ground state does not include
any states that can be identified as the {}``degenerate'' ground
states. Unfortunately, we can not definitively conclude whether this
is because the topological degeneracy for this system is absent or
the system sizes we can simulate are too small to see the {}``topologically
degenerate'' states. In particular, the correlation length may be
such that those states lie above the first excited state for the simulated
system sizes. 

\textbf{Background }In this work, we are interested in the Hubbard
model on a honeycomb lattice given by 
\[
H=-t\sum_{ij,\sigma}c_{i\sigma}^{\dagger}c_{j\sigma}+U\sum_{i}(n_{i\uparrow}+1/2)(n_{i\downarrow}-1/2)+\mu
\]
where $\sigma$ denotes spin $\uparrow,\downarrow$, $n_{i\sigma}=c_{i\sigma}^{\dagger}c_{i\sigma}$
and we set the chemical potential $\mu=0$ so the system is at half
filling. Interest in this system was stimulated by numerical evidence
\cite{meng2010quantum} for a spin liquid. Since this time, a number
of concrete spin liquid states have been proposed \cite{clark2010nature,lu2010spin,wang2010schwinger,xu2011quantum}.
These states are primarily $Z_{2}$ spin liquids and therefore have
(on a torus) four degenerate ground states. The latter spin liquid
proposal \cite{xu2011quantum} involves additional symmetries and
requires a 16-fold degeneracy. Our primary focus in this paper will
be searching for four degenerate ground states. 

\begin{figure}
\includegraphics[clip,width=0.8\columnwidth]{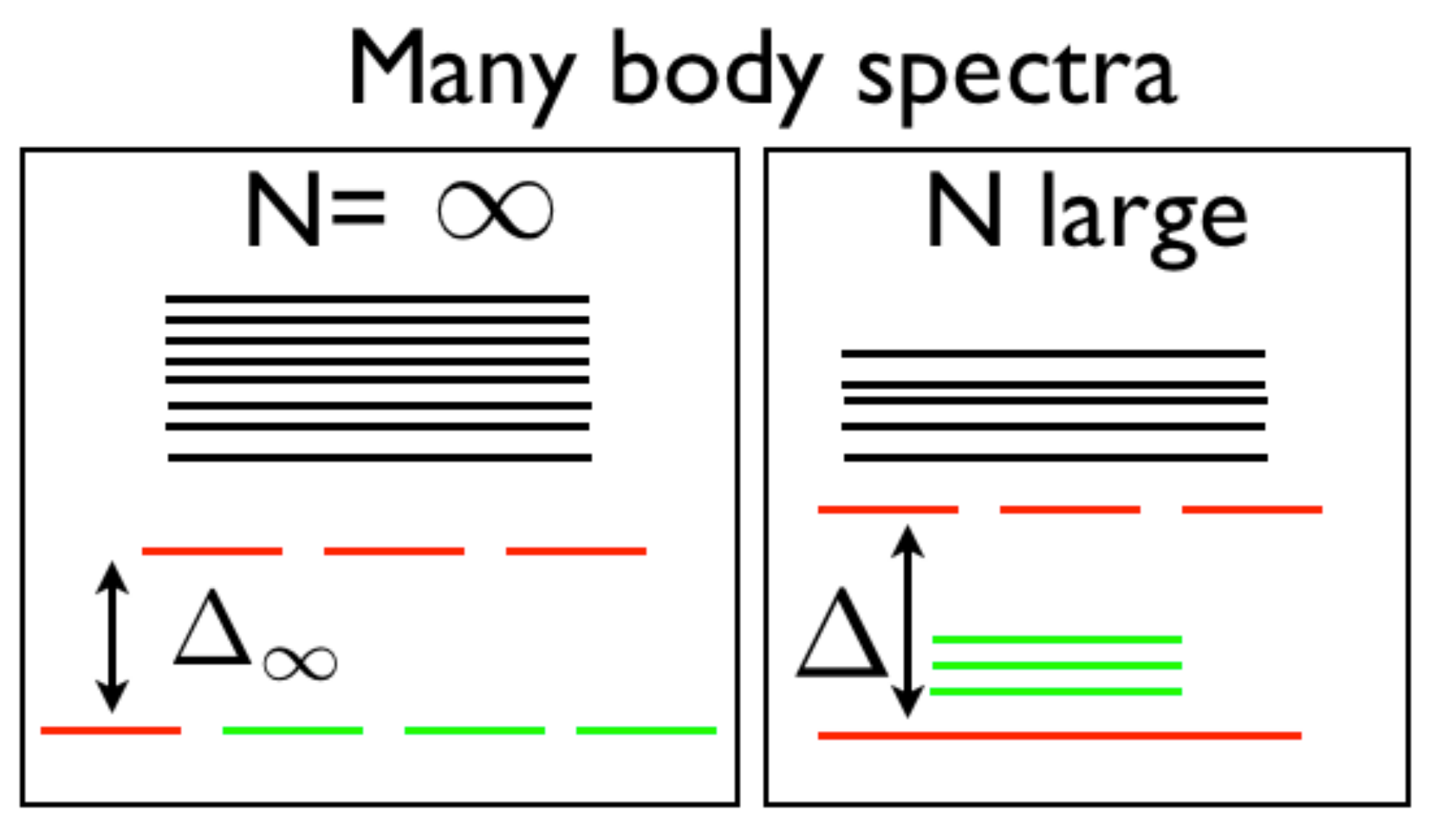}

\caption{\label{fig:Cartoon-representation-of}Cartoon representation of the
expected spectra for the $Z_{2}$ spin liquid on the Hubbard honeycomb
model for $N=\infty$ and large $N$. At $N=\infty$ (on a torus),
there are four degenerate ground states and then a gap to the (in
this case) three degenerate excited states. The four degenerate ground
states (shown in green) split at finite $N$ due to finite size effects;
we label them TDGS. The finite size-ground state and three excited
states (shown in red) have a (finite-size) gap identified by ref \cite{meng2010quantum};
these four states are labelled as the reference spectrum and used
as a baseline to compare against spectra which include TDGS. Other
excited states lie above the first excited state, but such states
can only increase the entropy $S(\beta).$ }
\end{figure}

In any finite system, the degeneracy of these states will be broken
leaving a single {}``finite-size ground state'' and three additional
{}``finite-size excited states'' in the spectrum (see fig. \ref{fig:Cartoon-representation-of}).
In the limit of large $N$, these states will approach the true ground
state exponentially quickly. Therefore evidence for topological degeneracy
could be found by looking for three excited states in the low-lying
spectrum that are not identifiable as other excitations (i.e. spin/charge
excitations) and ideally fall (significantly) below the true first
excited state. We denote these extra three states as thermodynamically
degenerate ground states (TDGS). The true thermodynamic first excited
state is identified by ref. \cite{meng2010quantum} as coming from
the staggered spin sector and therefore is presumably a three-fold
degenerate spin triplet (this fact is also consistent with our results).
We notate the ground state and three true first excited states as
the reference spectrum. 
 It is the existence of states (significantly) below this in which
we are therefore most interested. In this work, for finite size systems,
we (a) examine whether the available data can be accurately fit without
resorting to the existence of additional eigenstates in addition to
the reference spectrum (such as the TDGS) and (b) place bounds on
how low the TDGS can be in the many-body spectrum. Our focus will
be at $U=4$ where the system appears to have no broken symmetries. 

The methodology we use in this work is a finite temperature version
of Determinant Quantum Monte Carlo (FTDQMC) \cite{blankenbecler81}.
Using FTDQMC we can compute properties of eqn. (1) at finite temperature
which is naturally sensitive to all excitations independent of their
origin. This is to be contrasted to the approach used in ref. \cite{meng2010quantum}
which is a projection technique. This latter technique is designed
to project down to the true finite size ground state and, although
sensitive to the gap, this sensitivity strongly depends on the overlap
of the finite-size excited states with the initial trial state $\Psi_{T}$.
This means the TDGS might be (for all practical purposes) invisible
to this ground state method. 

\textbf{Entropy} One place where a ground state degeneracy should
be manifest is in the entropy at large inverse temperature $\beta\equiv1/k_{B}T$.
In particular, in the thermodynamic limit, a degenerate system will
have a ground state entropy of the logarithm of the degeneracy. In
a finite system, although this degeneracy breaks, one might expect
to see instead plateaus at large $\beta$. One nice feature of the
entropy is that it is a completely unbiased way to search for the
presence of excitations as it is sensitive only to their energy eigenvalues
placing no constraints on accurately identifying their quantum numbers,
etc. 

There are many approaches for computing the entropy of a physical
system. In this paper, we compute the entropy from many body spectra
as well as from the functional dependence of the energy $U(\beta)$
on the inverse temperature $\beta$.

Given a set of eigenvalues $E_{i}$ corresponding to a many body spectra,
the entropy of the system at a given $\beta$ is 
\[
S(\beta)=-\sum_{i}p_{i}(\beta)\ln p_{i}(\beta)
\]
 where $p_{i}$ is the probability of the eigenstate $i$, 
\[
p_{i}(\beta)=\frac{\exp(-\beta E_{i})}{\sum_{i}\exp(-\beta E_{i})}
\]
On the other hand the temperature dependence of $U(\beta)$ can be
used to compute the entropy via thermodynamic integration as 
\begin{equation}
S(\beta)=S(0)-\left[\int_{0}^{\beta}U\, d\beta-\beta U(\beta)\right]\label{eq:EntropyFromEnergy}
\end{equation}
As our simulations are run in the grand canonical ensemble, we always
have that $S(0)=4^{N}$ where $N$ is the total number of sites in
our system. 

\begin{figure}
\includegraphics[clip]{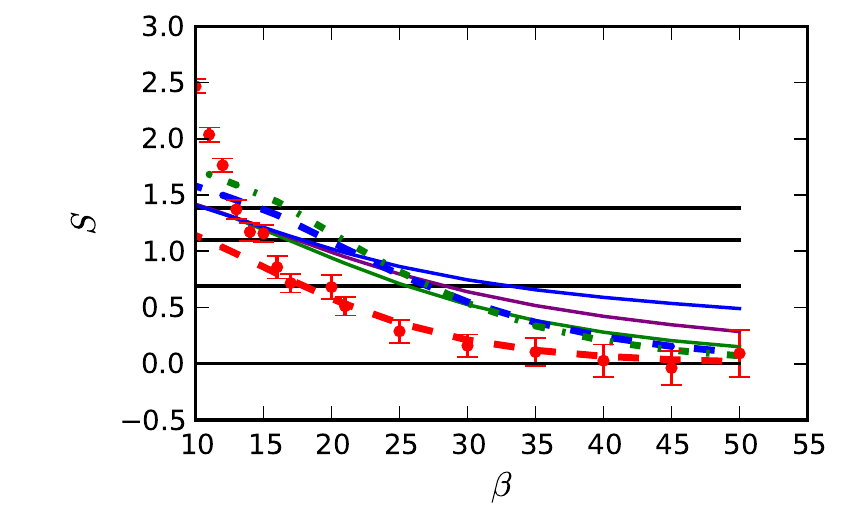}

\includegraphics[clip]{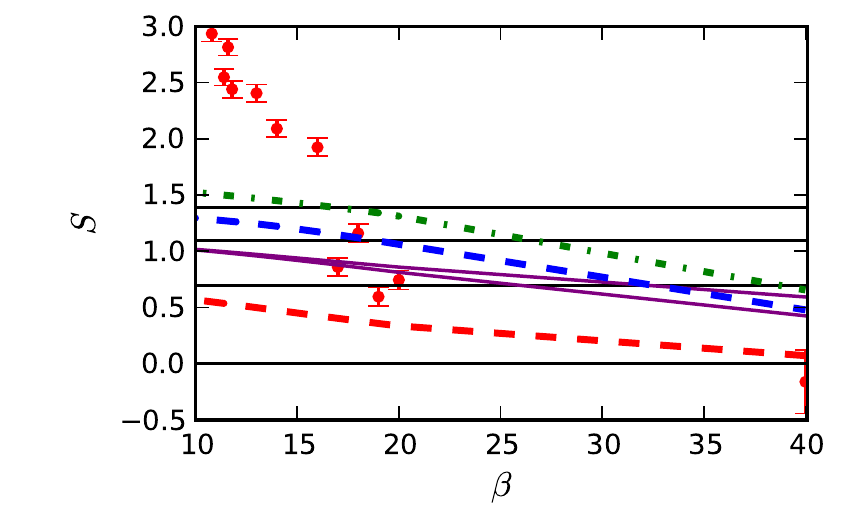}

\caption{\label{fig:Entropy}Entropy (red dots) for $N=72$ at $\tau=0.01$
(top) and $N=162$ at $\tau\rightarrow0$ (bottom). Time step errors
and discretization errors have been checked to be negligible. The
red dotted line shows the entropy of the reference spectrum using
$\Delta$ from \cite{meng2010quantum}. All other curved lines assume
this spectrum plus additional states. Curved solid lines (from top
to bottom) include an additional TDGS at (0.03,0.05,0.07) (top) and
(0.03,0.05) (bottom) respectively. The dashed blue line is the manifold
of lowest entropies for having two TDGS below 0.1 (top) or 0.07 (bottom).
The dot-dashed green line is the manifold of lowest entropies for
having three TDGS below 0.12 (top) or 0.07 (bottom). Black lines are
visual guides to the eye at 0, $\ln2,\ln3,\ln4$.}
\end{figure}

For each $\beta$ (on a fine grid) we compute the energy of the system.
This energy can then be converted to an entropy using eqn. \ref{eq:EntropyFromEnergy}
and seen in fig. \ref{fig:Entropy} for $N=72,162$. Notice that the
entropy does not plateau at $\log4$ (or $\log3$ or $\log2$) as
would be expected if three (or two or one respectively) TDGS lie significantly
below the finite-size first excited state. We additionally plot the
entropy for the reference spectrum using the gap $\Delta$ 
found in ref. \cite{meng2010quantum}
assuming a triply degenerate excited state. We find that this curve
favorably fits the entropy at large $\beta$ giving no indication
of the need for additional low lying eigenstates. Of course, a priori,
it is not implausible that a series of additional excitations above
and below the first excited state conspire to produce a similar entropy
at large $\beta$. To better quantify against this possibility, we
make use of the following fact (see supplemental information): Fix
the lowest $k$ excited states of a many-body spectrum. Any additional
higher-energy excited states can only make the entropy $S(\beta)$
increase for all $\beta.$ We proceed then to propose possible low
energy spectra, compute $S(\beta)$ associated with these spectra
and show that this entropy curve lies above the simulation data. As
additional high energy excited states can only increase the entropy,
this shows the data is inconsistent with the proposed low energy spectra.
We focus on low-energy spectra involving states in the reference spectrum
with the addition of 1-3 excitations below the true excited states.
Fig. \ref{fig:Entropy} (top) shows the entropy curves for a series
of low-energy spectra for $N=72$ that has a single additional TDGS
at energies ranging from half the first excited state to 0.1 as well
as the lowest entropy manifold for two states below 0.1 and three
states below 0.12. Simulations for $N=162$ are also shown.  Notice
the entropy of the proposed spectra all lie above the data ruling
out the possibility of a state at these energies. 

\begin{figure}
\includegraphics{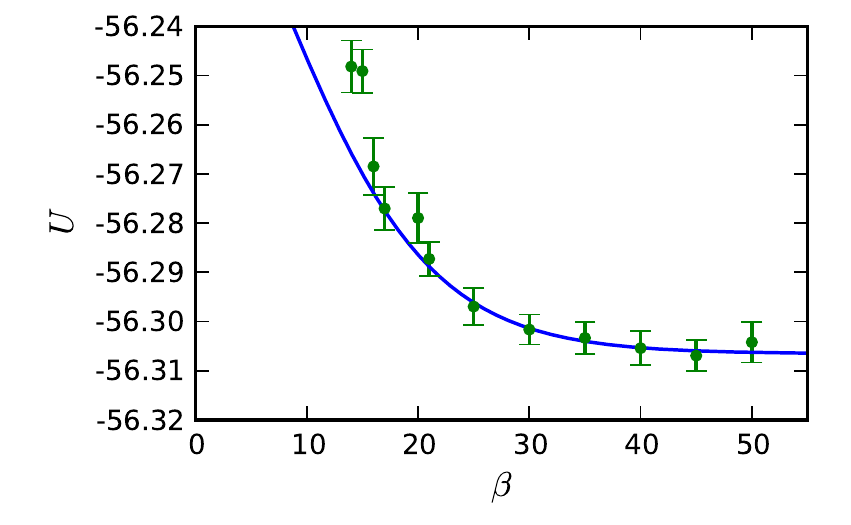}

\includegraphics{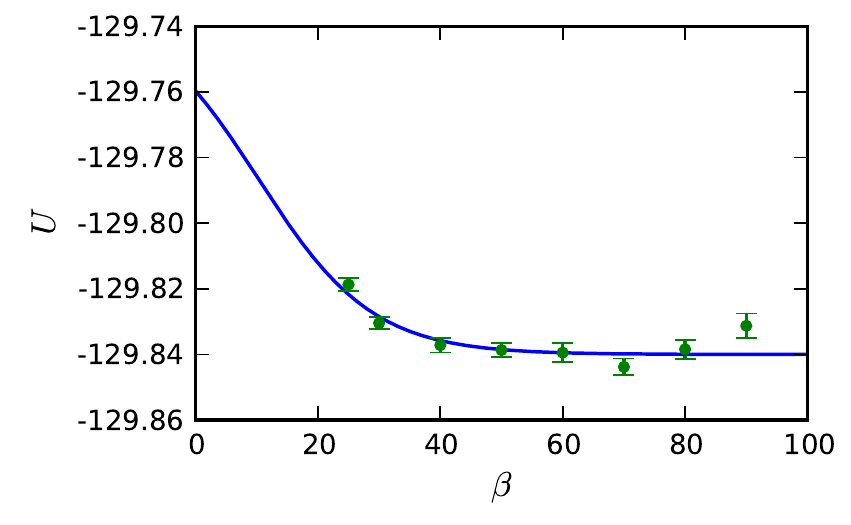}

\caption{Energy of $N=72$ (top) at $\tau=0.01$ and $N=162$ (bottom) at $\tau=0.1$.
The blue lines are $U(\beta)$ generated from the reference spectrum
with a gap $\Delta=(0.1469226,0.1071340)$ for $N=(72,162)$ taken
from ref. \cite{meng2010quantum} with a single free parameter for
the ground state energy selected to be $(-56.31,-129.84)$ respectively.
There are no indications that additional energies are needed to fit
$U$ at large $\beta$. \label{fig:Energy} }
\end{figure}

\textbf{Energy }A complementary approach to finding the eigenvalues
of the low-lying spectra is to fit it to the internal energy $U(\beta)$
at large $\beta$. In particular, we take the reference spectrum and
compute $U(\beta)=1/Z\sum_{i}E_{i}\exp[-\beta E_{i}]$ where $Z=\sum_{i}\exp[-\beta E_{i}]$.
It should be noted that while the entropy calculation (through thermodynamic
integration, eqn. \ref{eq:EntropyFromEnergy}) depends on the energies
at small $\beta$ ($\beta\lessapprox25$ for the $N=72$), fitting
$U(\beta)$ to a low-lying spectrum is sensitive to energies at large
$\beta$ ($\beta\gtrapprox25$) which is significantly below the gap.
In fig. \ref{fig:Energy} we plot the energy corresponding to the
reference spectrum for $N=72,162$ with the gap $\Delta$ found in
ref. \cite{meng2010quantum} assuming again a triply degenerate excited
state. We see that for both system sizes, the data is fit without
the need for additional low-energy states to be assumed. 

The nature of topologically distinct ground states is that there aren't
any local operators that convert one topological ground state to the
other. One, then, might be concerned that a quantum Monte Carlo approach
that samples space by applying the operator $\exp[-\tau H]$ (admittedly
in a auxiliary field space) might lock itself into one topological
sector, effectively becoming non-ergodic. We argue, that, although
possible, there is evidence that this is not the case. At infinite
temperature ($\beta=0$), we know that there are $4^{n}$ states which
correctly sets $S(0)$. In addition, even at high, but non-infinite
temperatures we suspect that the thermal fluctuations are sufficient
to `see' the topological states anticipating that if the topological
state is not sampled, it only happens at lower temperatures. Since
the entropy at $\beta$ depends only on energies at temperatures higher
then \textbf{$\beta$}, it would require the state to be affecting
the energies at temperatures higher then $\beta\approx15-20$ to affect
the conclusions of our study. Also, if the simulation is missing states
(at any temperature) this affects the energies at that temperature
and, by having an incorrect integrand, the entropy. The entropy would
then generically not integrate down to zero; our results are zero
within error bars and so any large deviation is inconsistent with
the data. An additional check of consistency is that the high energy
(i.e. entropy) data, the low energy data, and the spin gap measured
by ref. \cite{meng2010quantum} are all consistent with the same gap
$\Delta$; moreover there is no obvious jump or non-monotonic behavior
in the energies (which might be anticipated if some state disappeared
for some temperatures). Finally, over many simulations, we have seen
no evidence that the simulation sometimes finds one ground state and
sometimes a {}``higher energy one'' at low $T$ as might be expected
if it was randomly locking itself into one of many topological sectors.
Although these considerations do not completely rule out the presence
of non-ergodicity issues, we feel it makes it unlikely. 

\textbf{Discussion} In this work, we have shown that, up to $N=162$
sites, the evidence is strongly against the presence of states significantly
below the spin triplet. Although the finite-size splitting between
the TDGS are expected to decay exponentially with system size, there
is the mundane possibility that at the system sizes at which we are
able to perform calculations, the TDGS actually live high in the spectra
and would come down for larger sizes. Although always impossible to
rule out, the lack of any identifiable low lying states suggests  alternative
explanations. We suggested three alternate possibilities in the introduction
and find that there is supporting evidence in the literature for all
three possibilities. Recent work \cite{sorella2012absence,assaad2013pinning}
has reconsidered the problem and reached a different conclusion then
ref. \cite{meng2010quantum} about the absence of magnetic order at
$U\approx4$ arguing for the existence of an ordered phase. Alternatively,
ref. \cite{chang2012quantum} considered a related model, the staggered-flux
Hubbard model on a square lattice using an accurate (albeit approximate)
method and found evidence for a gapless spin liquid. As the staggered-flux
model can be continuously tuned to the honeycomb model by tuning the
nearest neighbor hoppings, this can be take as some evidence that
a spin liquid phase on the honeycomb is more likely to be gapless
then gapped; such a spin liquid would have no TDGS. Finally, there
has been recent interest in both extending Hastings theorem to rule
out featureless, non-topological states in other models as well as
explicit examples where featureless states can exist \cite{kimchi2012featureless}.
Although none of these results currently speak directly to the honeycomb
model, they do motivate the possibility that this is a featureless
state which is consistent with the entirety of the results of \cite{meng2010quantum}. 

Beyond giving evidence concerning the absence of the spin liquid phase
on the Hubbard model, we have also demonstrated an approach which
allows for identifying (or ruling out) the presence of degenerate
ground states using quantum Monte Carlo. In particular, we use the entropy
curve of model spectra as an effective way of showing that certain
low-lying spectra are inconsistent with data irrespective of the continuum
of excited states above them. Although not as accurate as exact diagonalization,
this allows calculations on systems sizes (for sign-free problems)
that go significantly beyond those that are reachable with a method
which scales exponentially and we hope that these results encourage
calculation of degenerate ground states in other putative spin liquid
candidates. \\

\textbf{Acknowledgements} We thank Fahker Assaad, Doron Bergman, David
Huse, Michael Kolodrubetz, Joseph Maciejko, David Pekker, Shivaji
Sondhi, and Ronny Thomale for helpful conversations as well as Bela
Bauer for helpful conversations and comments on this manuscript. This
work used the Extreme Science and Engineering Discovery Environment
(XSEDE), which is supported by National Science Foundation grant number
OCI-1053575. 

\bigskip{}

\appendix
\textbf{\uline{Supplemental information}}

\begin{flushleft}
\begin{figure}
(a) \includegraphics[scale=0.2]{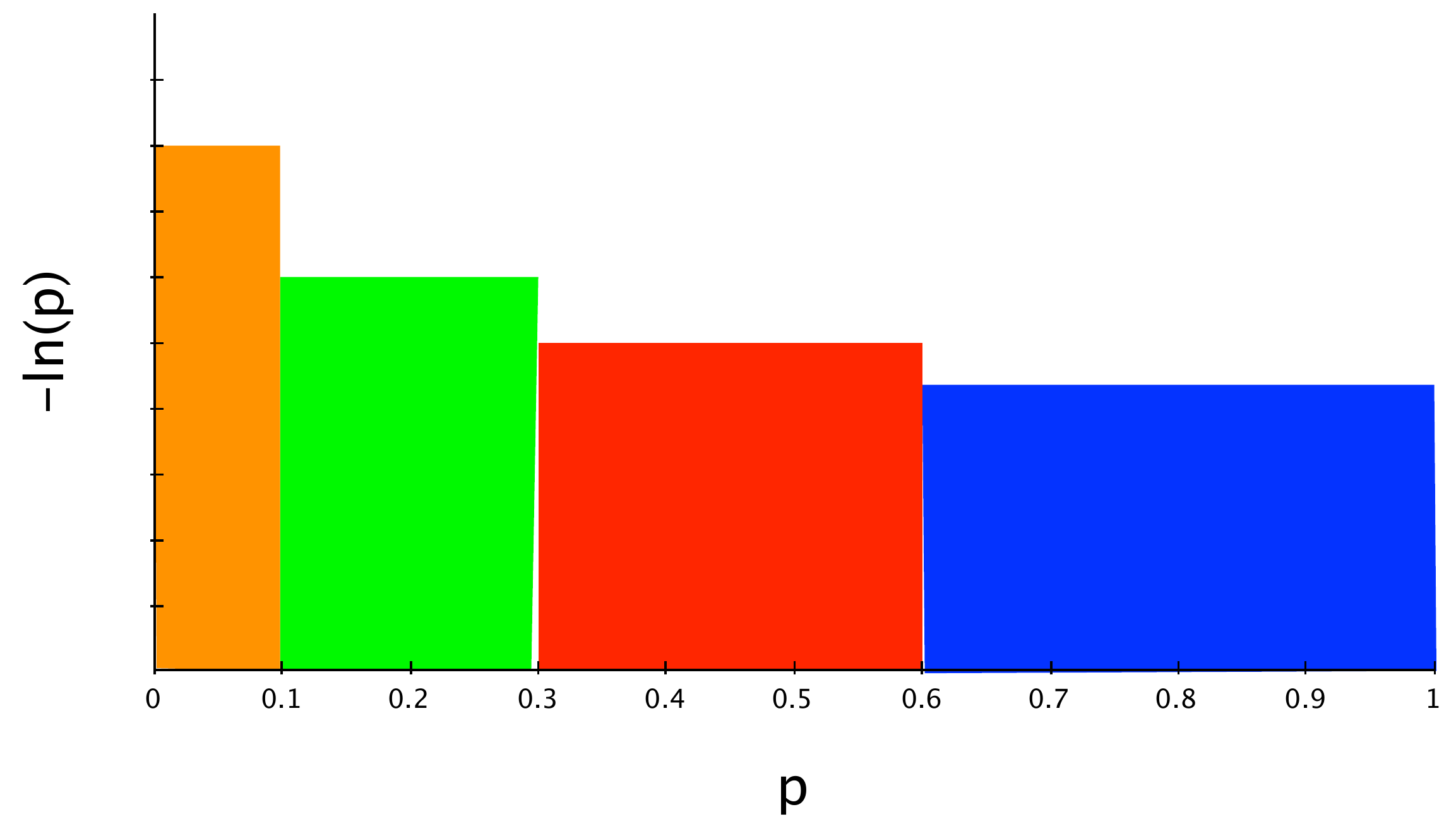}

(b) \includegraphics[scale=0.2]{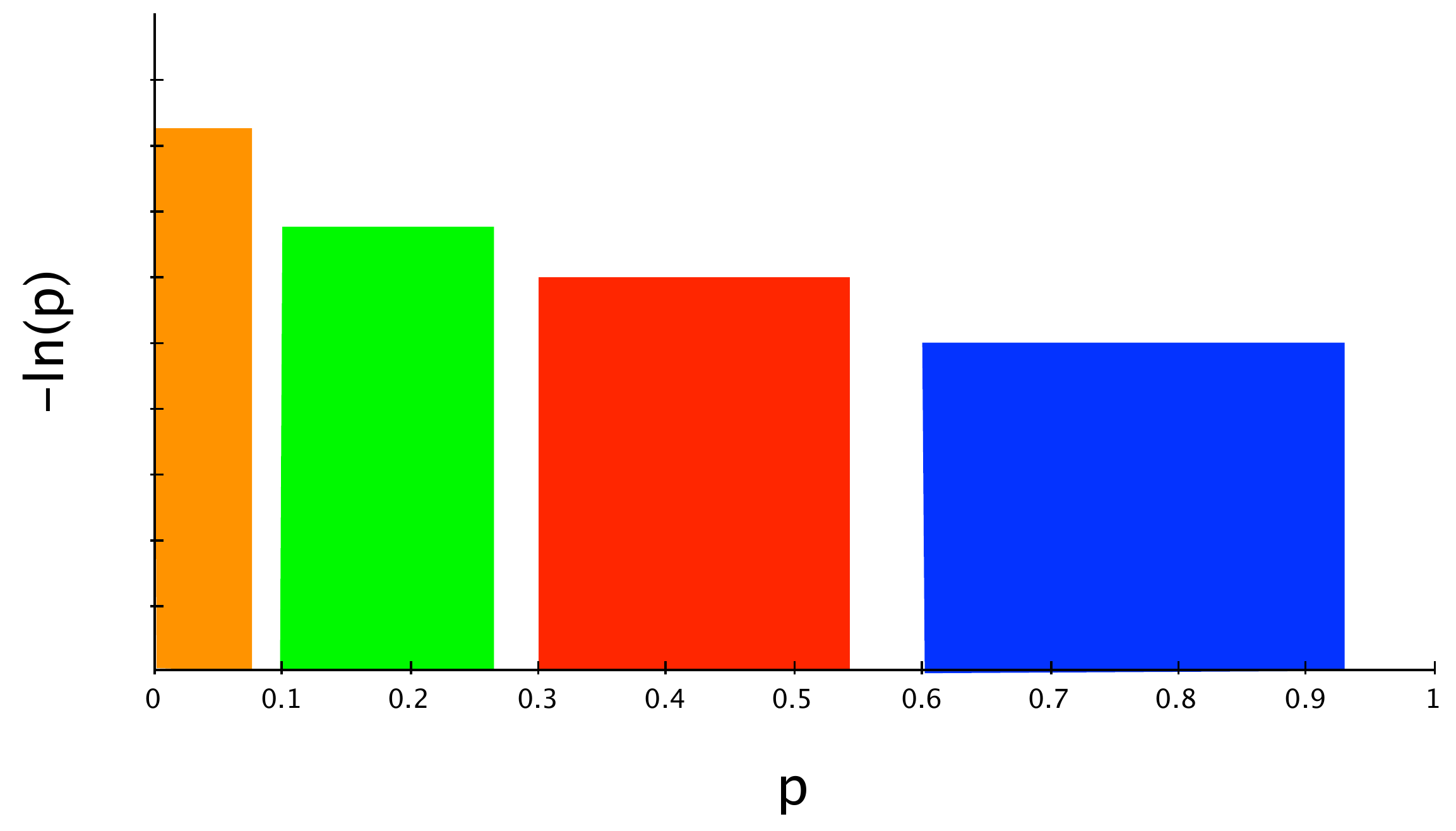}

(c) \includegraphics[scale=0.2]{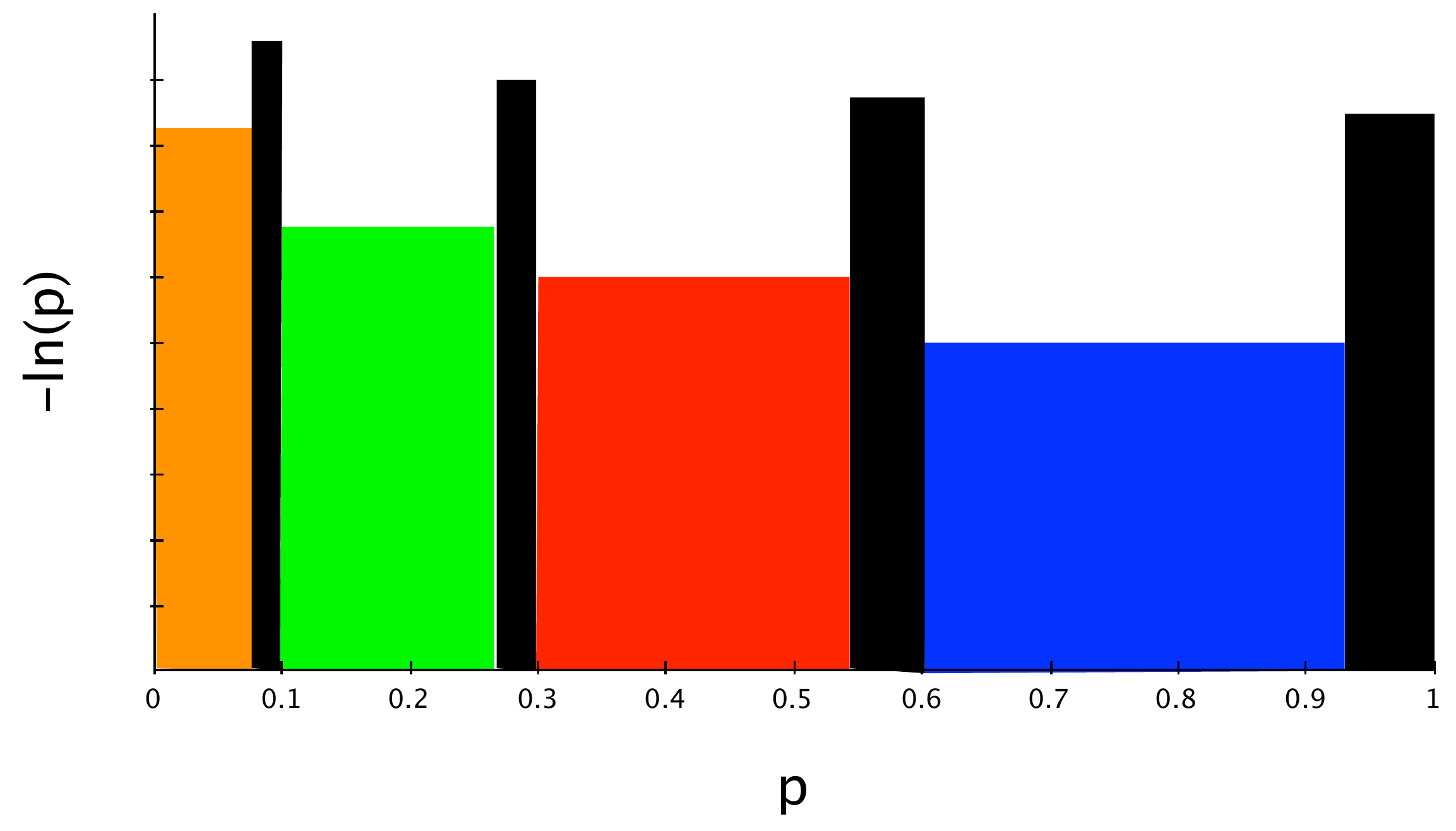}

\caption{Area under the figures represent the entropy of a series of states:
$-\sum_{i}p_{i}\ln p_{i}$. As more states are added which have smaller
probability of being sampled, the area must increase. \label{fig:EntropyProof} }
\end{figure}
 In this appendix, we give a short proof that adding additional states
in a spectrum above the largest energy $E_{min}$ of a fixed low-energy
reference spectrum can only increase the entropy. 
\par\end{flushleft}

\begin{flushleft}
Although there are many way to show this fact, we use the following
geometric argument. The entropy is defined as $S=-\sum_{i}p_{i}\ln p_{i}$
. We can represent this geometrically as the area in a chart prototypically
shown in fig. \ref{fig:EntropyProof}(a). Notice that we have $\sum_{i}p_{i}=1$
where each $p_{i}=\exp[-\beta E_{i}]/Z$ with $Z=\sum_{i}\exp[-\beta E_{i}]$.
When we add additional states the probability of the original states
decreases as $p_{i}=\exp[-\beta E_{i}]/\tilde{Z}$ where $\tilde{Z}>Z$
as the sum is over more states. As $-\ln p$ monotonically increases
as $p$ gets smaller , the {}``height'' taken by each of the former
states in the reference spectrum strictly increases as see in fig.
\ref{fig:EntropyProof}(b). By assumption, the added states are above
all the original states in the reference spectrum (i.e. $E_{new}>E_{old}$)
and therefore, for each $j$ in the added states, $p_{j}=\exp[-\beta E_{j}]/\tilde{Z}$
must be smaller then each $p_{i}=\exp[-\beta E_{i}]/\tilde{Z}$ in
the reference spectrum. Therefore, the {}``height'' of each additional
state must be larger then the height of any state in the reference
spectrum as seen in fig. \ref{fig:EntropyProof}(c). Therefore, the
total entropy (the area in the figure) must go up. 
\par\end{flushleft}

\bibliographystyle{plainnat}
\bibliography{FreeEnergy}

\end{document}